\documentclass{INTERSPEECH2023}

\interspeechcameraready

\usepackage{cite}
\PassOptionsToPackage{hyphens}{url}
\usepackage{hyperref}

\title{MP-SENet: A Speech Enhancement Model with Parallel Denoising of Magnitude and Phase Spectra}
\name{Ye-Xin Lu, Yang Ai$^*$, Zhen-Hua Ling\thanks{$^*$ Corresponding author. This work was partially funded by the Fundamental Research Funds for the Central Universities.}}
\address{
    National Engineering Research Center of Speech and Language Information Processing, \\University of Science and Technology of China, Hefei, P. R. China}

\email{yxlu0102@mail.ustc.edu.cn, \{yangai, zhling\}@ustc.edu.cn}

\begin{document}

\maketitle

\begin{abstract}
This paper proposes MP-SENet, a novel Speech Enhancement Network which directly denoises Magnitude and Phase spectra in parallel. The proposed MP-SENet adopts a codec architecture in which the encoder and decoder are bridged by convolution-augmented transformers. The encoder aims to encode time-frequency representations from the input noisy magnitude and phase spectra. The decoder is composed of parallel magnitude mask decoder and phase decoder, directly recovering clean magnitude spectra and clean-wrapped phase spectra by incorporating learnable sigmoid activation and parallel phase estimation architecture, respectively. Multi-level losses defined on magnitude spectra, phase spectra, short-time complex spectra, and time-domain waveforms are used to train the MP-SENet model jointly. Experimental results show that our proposed MP-SENet achieves a PESQ of 3.50 on the public VoiceBank+DEMAND dataset and outperforms existing advanced speech enhancement methods.
\end{abstract}


\section{Introduction}
In real-life scenarios, speech waveforms captured by devices are inevitably degraded by noises, which immensely impacts real applications such as hearing aids and telecommunications. To alleviate the impact of noises and improve speech perceptual quality, many deep-learning-based speech enhancement (SE) methods have been proposed to recover clean waveforms from the degraded ones utilizing neural networks. Existing SE methods can be roughly divided into two categories, i.e., time-domain SE methods and time-frequency (TF) domain SE methods. The time-domain SE methods \cite{pascual2017segan, pandey2019tcnn, defossez2020real, kim2021se, kong2022speech} adopted neural networks to learn the mapping from noisy waveforms to clean ones. Unfortunately, this category of methods still suffered from quality bottlenecks and showed inefficiency due to the direct generation of high-resolution waveforms. In comparison, the TF-domain SE methods exhibited superior performance.

The TF-domain SE methods aim to predict clean frame-level TF-domain representations and then reconstruct the enhanced waveforms. Generally, the phase is not included in the commonly used representations because it is a great challenge to directly enhance the phase spectra, given its wrapping and nonstructural properties. However, recent studies demonstrated that the phase information plays an essential role in the speech perceptual quality of SE methods, especially in low signal-to-noise (SNR) circumstances \cite{paliwal2011importance}. In earlier studies, researchers only enhanced the magnitude spectra and reconstructed the waveforms using inverse short-time Fourier transform (ISTFT) from the enhanced magnitude and noisy phase spectra \cite{valentini2018speech,ai2019dnn,xu2014regression,kim2020t}. The absence of phase spectrum enhancement inevitably led to the degradation of the enhanced speech quality.
To overcome the above issues, several approaches focused on the enhancement of short-time complex spectra, which implicitly recovered both clean magnitude and phase spectra \cite{tan2019learning, dang2022dpt, yin2022tridentse}. More recent studies also proposed to enhance the magnitude followed by complex spectrum refinement \cite{yu2022dual, cao2022cmgan}, which can alleviate the unbounded estimation problem  \cite{williamson2015complex} existing in the methods that only enhanced complex spectra. However, the compensation effect \cite{wang2021compensation} between the magnitude and phase still existed, which led to imprecise phase estimation. Although PHASEN \cite{yin2020phasen} proposed a two-stream network and acquired the ability to handle detailed phase patterns and utilize harmonic patterns, it still optimized the phase within the complex spectrum level. These methods were still unable to precisely and explicitly predict the clean phase spectra, leaving room for improvement in the enhanced speech quality. To this end, it is crucial to implement explicit prediction and optimization on the phase spectra for TF-domain SE methods.

Therefore, we propose MP-SENet, a TF-domain monaural SE model with parallel magnitude and phase spectra denoising. The MP-SENet forms a codec architecture and we bridge the encoder and decoder using two-stage convolution-augmented transformers (TS-Conformers) borrowed from CMGAN \cite{cao2022cmgan} to capture both local and global information.
The encoder encodes the input noisy magnitude and phase spectra to compressed TF-domain representations for subsequent decoding. The parallel magnitude mask decoder and phase decoder decode out the clean magnitude and phase spectra, respectively, and finally the enhanced waveforms are reconstructed by ISTFT. Specifically, the magnitude mask decoder predicts boundary masks with a learnable sigmoid activation and then multiplies the masks with the noisy magnitude spectra to obtain the clean magnitude spectra. Inspired by our previous work \cite{ai2022neural}, the phase decoder incorporates the parallel phase estimation architecture to directly predict the clean phase spectra. Experimental results show that our proposed MP-SENet outperforms state-of-the-art (SOTA) SE methods and alleviates the compensation effect between the magnitude and phase spectra by achieving explicit predictions and optimizations of them. To the best of our knowledge, we are the first to accomplish the direct enhancement of phase spectra.

\begin{figure*}[htbp!]
  \centering
  \includegraphics[width=\linewidth]{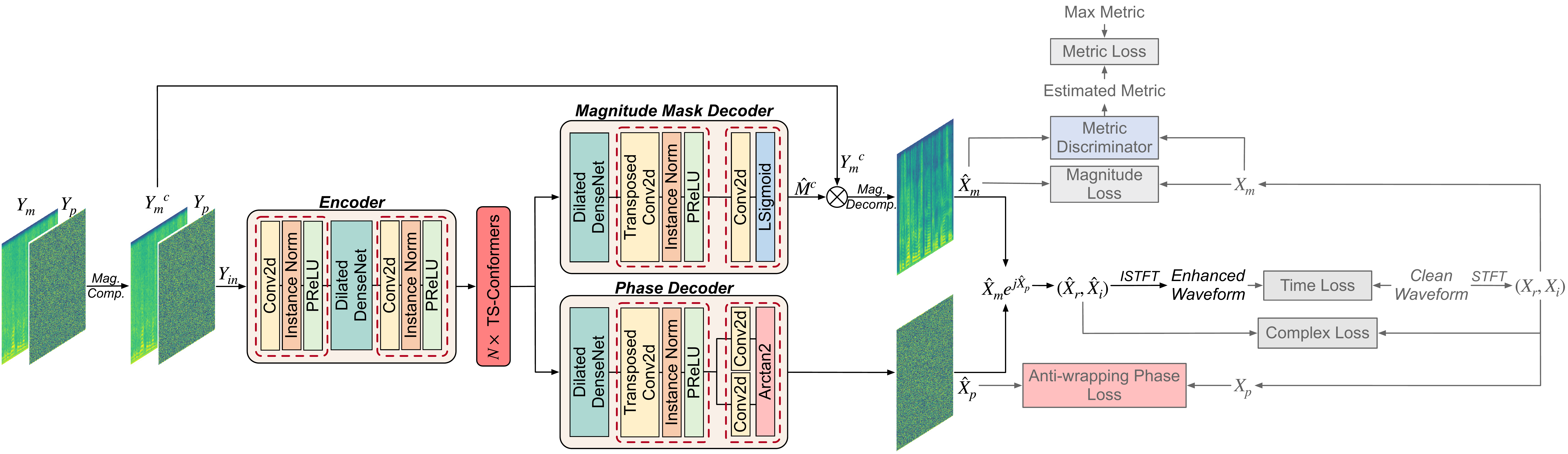}
  \caption{Overall structure and training principles of the proposed MP-SENet. The ``Mag. Comp." and ``Mag. Decomp." denote the magnitude compression and magnitude decompression operations, respectively. The dim parts only appear at the training stage.}
  \label{fig: model}
\end{figure*}

\section{Methodology}
The overview of the model structure and training criteria of the proposed MP-SENet are illustrated in Fig.~\ref{fig: model}. The MP-SENet adopts a codec architecture to denoise the noisy speech waveform $\bm{y} \in \mathbb{R}^L$ and recover the clean speech waveform $\bm{x} \in \mathbb{R}^L$ in the TF domain, where $L$ is the waveform length.
Specifically, we first extract the magnitude spectrum $\bm{Y_m} \in \mathbb{R}^{T\times F}$ and the wrapped phase spectrum $\bm{Y_p} \in \mathbb{R}^{T\times F}$ from $\bm{y}$ through STFT, where $T$ and $F$ denote the total number of frames and frequency bins, respectively. For more precise magnitude mask prediction, we apply the power-law compression on $\bm{Y_m}$ and stack it with $\bm{Y_p}$ to compose an input feature $\bm{Y_{in}} \in \mathbb{R}^{T \times F \times 2}$. Then the encoder encodes $\bm{Y_{in}}$ into a compressed TF-domain representation, and subsequently, the TF-domain representation is processed by four TS-Conformers to capture time and frequency dependencies stage by stage. Finally, the parallel magnitude mask decoder and phase decoder predict the clean magnitude spectrum $\bm{\hat{X}_m}$ and clean phase spectrum $\bm{\hat{X}_p}$ from the TF-domain representation, respectively, and the final enhanced waveform $\bm{\hat{x}}$ is reconstructed by ISTFT. The details of the encoder, magnitude mask decoder, phase decoder, and training criteria are described as follows.

\vspace{-1mm}
\subsection{Model structure}
\vspace{-1mm}

\vspace{-1mm}
\subsubsection{Encoder}
\vspace{-1mm}

As illustrated in Fig.~\ref{fig: model}, the encoder encodes the input feature $\bm{Y_{in}}$ into a TF-domain representation with a lower sampling rate and higher dimensions. It is a cascade of a convolutional block, a dilated DenseNet \cite{pandey2020densely}, and another convolutional block. Each convolutional block consists of a 2D convolutional layer, an instance normalization \cite{ulyanov2016instance}, and a parametric rectified linear unit (PReLU) activation \cite{he2015delving}. The first convolutional block increases the feature dimension by increasing the number of channels in the convolutional layer, while the second block downsamples the feature by expanding the stride in the convolutional layer.
The dilated DenseNet utilizes four convolutional layers with dilation sizes of ${1, 2, 4, 8}$ to extend the receptive field along the time axis, and applies dense connections to all the convolutional layers to avoid the vanishing gradient problem. 

\vspace{-1mm}
\subsubsection{Magnitude mask decoder}
\vspace{-1mm}

As illustrated in Fig.~\ref{fig: model}, the magnitude mask decoder predicts a magnitude mask from the TF-domain representation and multiplies it with the noisy magnitude spectrum to obtain the clean magnitude spectrum. However, the commonly used magnitude mask $\bm{M} = \bm{X_m} / \bm{Y_m}  \in \mathbb{R}^{T \times F}$ is unbounded, where $\bm{X_m}$ denotes the magnitude spectrum of the clean waveform $\bm{x}$. Earlier studies usually employ the sigmoid function to limit the value range of the predicted mask $\bm{\hat{M}}$ to $(0,1)$. There is an insurmountable gap between the predicted mask and the real mask whose range is out of $(0, 1)$. Recent methods \cite{yu2022dual, cao2022cmgan} compensate the gap by introducing another stream of complex spectrum prediction. Nevertheless, the compensation effect between magnitude and phase spectra still exists, leading to discontinuity of the spectrogram and damage to its harmonic structure. Accordingly, we first apply power-law compression on $\bm{X_m}$ and $\bm{Y_m}$ with a compression factor $c$ to narrow the scope of the mask for easier prediction. Hence, the prediction target of the magnitude mask decoder is the compressed mask $\bm{M}^c = (\bm{X_m} / \bm{Y_m})^c $ and we set $c$ to 0.3 in the experiments. Subsequently, to achieve precise prediction, we further adopt the learnable sigmoid (LSigmoid) function from \cite{fu2021metricgan+} to predict the compressed magnitude mask:
\begin{equation}
	{\rm LSigmoid}(t) = \frac{\beta}{1+e^{1-\alpha t}},
\end{equation}
where $\beta$ is set to 2.0 and $\alpha \in \mathbb{R}^F$ is a trainable parameter, which allows the model to adaptively change the shape of the activation function in different frequency bands.

Specifically, with a TF-domain representation processed by TS-Conformers as input, the magnitude decoder feeds it to a dilated DenseNet, a deconvolutional block, and a magnitude mask estimation architecture to get the estimated compressed mask $\bm{\hat{M}}^c$. The deconvolutional block is used for upsampling and is composed of a 2D transposed convolutional layer, an instance normalization, and a PReLU activation. 
The magnitude mask estimation architecture first uses a 2D convolutional layer to reduce the output channel numbers of the deconvolutional block to 1 and then outputs ${\bm{\hat{M}}}^c$ with the activation of the LSigmoid function. Finally, the enhanced magnitude spectrum $\bm{\hat{X}_m}$ can be obtained by mask decoding as follows:
\begin{equation}
	\bm{\hat{X}_m} = ({\bm{Y_m}}^c \odot {\bm{\hat{M}}}^c)^{\frac{1}{c}},
\end{equation}
where $\odot$ denotes the element-wise multiplication.

\vspace{-1mm}
\subsubsection{Phase decoder}
\vspace{-1mm}

As illustrated in Fig.~\ref{fig: model}, the phase decoder directly predicts the clean phase spectrum from the TF-domain representation. In order to overcome the difficulties caused by the nonstructural and wrapping characteristics of the phase, we follow our previous work \cite{ai2022neural}, cascade a parallel phase estimation architecture after a dilated DenseNet and a deconvolutional block in the phase decoder. The parallel phase estimation architecture first adopts two parallel 2D convolutional layers to output the pseudo-real part component $\bm{\hat{X}_p^{(r)}}$ and pseudo-imaginary part component $\bm{\hat{X}_p^{(i)}}$, and then activates these two components to predict the clean wrapped phase spectrum $\bm{\hat{X}_p}$ using the two-argument arctangent (Arctan2) function, i.e.,

\begin{equation}
    \setlength{\abovedisplayskip}{3pt}
    \setlength{\belowdisplayskip}{3pt}
	\bm{\hat{X}_p} = \arctan\bigg(\frac{\bm{\hat{X}_p^{(i)}}}{\bm{\hat{X}_p^{(r)}}}\bigg) - \frac{\pi}{2} \cdot {\rm Sgn}^*(\bm{\hat{X}_p^{(i)}}) \cdot [{\rm Sgn}^*(\bm{\hat{X}_p^{(r)}}) - 1],
\end{equation}
where ${\rm Sgn}^*(t)$ is a redefined function which equals to 1 when $t \geq 0$, and equals to -1 when $t < 0$.

\vspace{-1mm}
\subsection{Training criteria}
\vspace{-1mm}

We define multi-level loss functions for training the proposed MP-SENet. In keeping with \cite{cao2022cmgan}, we use the time loss $\mathcal{L}_{\rm Time}$,  magnitude loss $\mathcal{L}_{\rm Mag.}$, and complex loss $\mathcal{L}_{\rm Com.}$, i.e.,
\begin{align}
	&\mathcal{L}_{\rm Time} = \mathbb{E}_{\bm{x}, \bm{\hat{x}}} [\Vert \bm{x} - \bm{\hat{x}}\Vert_1], \\
	&\mathcal{L}_{\rm Mag.} = \mathbb{E}_{\bm{X_m}, \bm{\hat{X}_m}} [\Vert \bm{X_m} - \bm{\hat{X}_m} \Vert_2^2], \\
	&\mathcal{L}_{\rm Com.}   = \mathbb{E}_{\bm{X_r}, \bm{\hat{X}_r}} [\Vert \bm{X_r} - \bm{\hat{X}_r} \Vert_2^2] + \mathbb{E}_{\bm{X_i}, \bm{\hat{X}_i}} [\Vert \bm{X_i} - \bm{\hat{X}_i} \Vert_2^2],
\end{align}
where $(\bm{X_r}, \bm{X_i}), (\bm{\hat{X}_r}, \bm{\hat{X}_i})$ denote the real and imaginary parts of the clean and enhanced complex spectra. Additionally, inspired by \cite{fu2019metricgan}, a metric discriminator is also adopted for generative adversarial training, and the perceptual evaluation of speech quality (PESQ) is used as the target metric. We rescale the PESQ score to $(0, 1)$, so the discriminator is trained to output the scaled PESQ score with pairs of clean and estimated magnitude spectra as inputs. The generator aims to output the magnitude spectrum whose metric can be discriminated to approach 1. The discriminator loss $\mathcal{L}_{\rm D}$ and the corresponding generated metric loss $L_{\rm Metric}$ are described as follows:
\begin{equation}
    \setlength{\abovedisplayskip}{3pt}
    \setlength{\belowdisplayskip}{3pt}
	\begin{aligned}
	\mathcal{L}_{\rm D} &= \mathbb{E}_{\bm{X_m}} [\Vert D(\bm{X_m}, \bm{X_m})-1 \Vert_2^2] \\
						&+ \mathbb{E}_{\bm{X_m}, \bm{\hat{X}_m}} [\Vert D(\bm{X_m}, \bm{\hat{X}_m})- Q_{\rm PESQ} \Vert_2^2],
	\end{aligned}
\end{equation}
\begin{equation}
    \setlength{\abovedisplayskip}{3pt}
    \setlength{\belowdisplayskip}{3pt}
	\mathcal{L}_{\rm Metric} = \mathbb{E}_{\bm{X_m}, \bm{\hat{X}_m}} [\Vert D(\bm{X_m}, \bm{\hat{X}_m})-1 \Vert_2^2],
\end{equation}
where $D$ denotes the discriminator and $Q_{\rm PESQ} \in [0,1]$ denotes the scaled PESQ score.

In previous TF-domain SE works, phase spectra are optimized within the complex spectra or by directly calculating the absolute $L^p$ distance between the natural clean and enhanced phase spectra $\bm{X_p}$ and $\bm{\hat{X}_p}$. However, due to the phase wrapping property, the absolute distance between two phases may not be their actual distance, revealing the inappropriateness of conventional losses (e.g., absolute $L^p$ distance) for phase optimization. Consistent with the anti-wrapping losses we proposed in \cite{ai2022neural}, we respectively define the instantaneous phase loss, group delay loss, and instantaneous angular frequency loss between $\bm{X_p}$ and $\bm{\hat{X}_p}$ as follows:
\begin{align}
    \setlength{\abovedisplayskip}{3pt}
    \setlength{\belowdisplayskip}{3pt}
    \mathcal{L}_{\rm IP} &= \mathbb{E}_{\bm{X_p}, \bm{\hat{X}_p}} [\Vert f_{\rm AW}(\bm{X_p} - \bm{\hat{X}_p}) \Vert_1], \\
	\mathcal{L}_{\rm GD} &= \mathbb{E}_{\Delta_{\rm DF} (\bm{X_p}, \bm{\hat{X}_p})} [\Vert f_{\rm AW}(\Delta_{\rm DF}(\bm{X_p} - \bm{\hat{X}_p})) \Vert_1], \\
	\mathcal{L}_{\rm IAF} &= \mathbb{E}_{\Delta_{\rm DT} (\bm{X_p}, \bm{\hat{X}_p})} [\Vert f_{\rm AW}(\Delta_{\rm DT}(\bm{X_p} - \bm{\hat{X}_p})) \Vert_1],
\end{align}
where $f_{\rm AW}(t) = \vert t - 2\pi \cdot {\rm round} \big(\frac{t}{2\pi}\big) \vert, t \in \mathbb{R}$ is an anti-wrapping function, which is used to avoid the error expansion issue caused by phase wrapping. $\Delta_{DF}$ and $\Delta_{DT}$ represent the differential operators along the frequency axis and time axis, respectively. The anti-wrapping phase loss is defined as:
\begin{equation}
    \setlength{\abovedisplayskip}{3pt}
    \setlength{\belowdisplayskip}{3pt}
	\mathcal{L}_{\rm Pha.} = \mathcal{L}_{\rm IP} + \mathcal{L}_{\rm GD} + \mathcal{L}_{\rm IAF}.
\end{equation}

The final generator loss $\mathcal{L}_{\rm G}$ is the linear combination of the losses mentioned above:
\begin{equation}
    \setlength{\abovedisplayskip}{3pt}
    \setlength{\belowdisplayskip}{3pt}
	\mathcal{L}_{\rm G} = \gamma_1\mathcal{L}_{\rm Time} + \gamma_2\mathcal{L}_{\rm Mag.} + \gamma_3\mathcal{L}_{\rm Com.} + \gamma_4\mathcal{L}_{\rm Metric} + \gamma_5\mathcal{L}_{\rm Pha.}.
\end{equation}
where $\gamma_1, \gamma_2,..., \gamma_5$ are hyperparameters and are set to 0.2, 0.9, 0.1, 0.05, and 0.3 with empirical trials, respectively. The training criteria of the MP-SENet is to minimize $\mathcal{L}_{\rm G}$ and $\mathcal{L}_{\rm D}$ simultaneously.

\begin{table*}[htbp!]
  \caption{Comparison with other methods on VoiceBank+DEMAND dataset. ``-" denotes the result is not provided in the original paper.}
  \label{tab: objective evaluation}
  \centering
  \begin{tabular}{lclccccccc}
    \toprule
    \textbf{Method} & \textbf{Year} & \textbf{Input} & \textbf{\#Param.} & \textbf{PESQ} & \textbf{CSIG} & \textbf{CBAK} & \textbf{COVL} & \textbf{SSNR} & \textbf{STOI} \\
    \midrule
    Noisy          					   & -     & -                  & -     & 1.97 & 3.35 & 2.44 & 2.63 & 1.68  & 0.91 \\
    \midrule
    SEGAN\cite{pascual2017segan}       & 2017  & Waveform           & 43.18M & 2.16 & 3.48 & 2.94 & 2.80 & 7.73  & 0.92 \\
    DEMUCS \cite{defossez2020real}     & 2021  & Waveform           & 33.53M   & 3.07 & 4.31 & 3.40 & 3.63 & -     & 0.95 \\
    SE-Conformer \cite{kim2021se}      & 2021  & Waveform           & -     & 3.13 & 4.45 & 3.55 & 3.82 & -     & 0.95 \\
    MetricGAN \cite{fu2019metricgan}   & 2019  & Magnitude          & -     & 2.86 & 3.99 & 3.18 & 3.42 & -     & -    \\
    MetricGAN+ \cite{fu2021metricgan+} & 2021  & Magnitude          & -     & 3.15 & 4.14 & 3.16 & 3.64 & -     & -    \\
    DPT-FSNet  \cite{dang2022dpt}      & 2021  & Complex            & 0.88M  & 3.33 & 4.58 & 3.72 & 4.00 & -     & \textbf{0.96} \\
    TridentSE \cite{yin2022tridentse}  & 2023  & Complex            & 3.03M  & 3.47 & 4.70 & 3.81 & 4.10 & -     & \textbf{0.96} \\
    DB-AIAT \cite{yu2022dual}          & 2021  & Magnitude+Complex  & 2.81M  & 3.31 & 4.61 & 3.75 & 3.96 & 10.79 & -    \\
    CMGAN \cite{cao2022cmgan}          & 2022  & Magnitude+Complex  & 1.83M  & 3.41 & 4.63 & 3.94 & 4.12 & \textbf{11.10} & \textbf{0.96} \\
    PHASEN \cite{yin2020phasen}        & 2020  & Magnitude+Phase    & - & 2.99 & 4.21 & 3.55 & 3.62 & 10.18 & -    \\
    \textbf{MP-SENet} & 2023  & Magnitude+Phase & 2.05M & \textbf{3.50} & \textbf{4.73} & \textbf{3.95} & \textbf{4.22} & 10.64 & \textbf{0.96} \\
    \bottomrule
  \end{tabular}
\end{table*}

\section{Experiments}
\subsection{Datasets and experimental setup}
\vspace{-1mm}
We used the publicly available VoiceBank+DEMAND dataset \cite{valentini2016investigating} for our experiments, which includes pairs of clean and noisy audio clips with a sampling rate of 48 kHz. The clean audio set is selected from the Voice Bank corpus \cite{veaux2013voice}, which consists of 11,572 audio clips from 28 speakers for training and 872 audio clips from 2 unseen speakers for testing. The clean audio clips are mixed with 10 types of noise (8 types from the DEMAND database \cite{thiemann2013diverse} and 2 artificial types) at SNRs of 0dB, 5dB, 10dB, and 15 dB for the training set and 5 types of unseen noise from the DEMAND database at SNRs of 2.5 dB, 7.5dB, 12.5 dB, and 17.5 dB for the test set.

We resampled all the audio clips to 16 kHz in the experiments. To extract input features from raw waveforms using STFT, the FFT point number, Hanning window size, and hop size were set to 400, 400 (25 ms), and 100 (6.25 ms), respectively. All the models were trained using the AdamW optimizer \cite{loshchilov2017decoupled} until 100 epochs. 
The learning rate was set initially to 0.0005 and halved every 30 epochs \footnote{Audio samples and source codes of the MP-SENet are available at \href{https://github.com/yxlu-0102/MP-SENet}{https://github.com/yxlu-0102/MP-SENet}.}.

\vspace{-1mm}
\subsection{Comparison with advanced SE methods}
\vspace{-1mm}

Several representative time-domain SE methods including SEGAN \cite{pascual2017segan}, DEMUCS \cite{defossez2020real} and SE-Conformer \cite{kim2021se}, and TF-domain SE methods including MetricGAN\cite{fu2019metricgan}, PHASEN \cite{yin2020phasen}, MetricGAN+ \cite{fu2021metricgan+}, and four SOTA methods (i.e., DPT-FSNet \cite{dang2022dpt}, TridentSE \cite{yin2022tridentse}, DB-AIAT \cite{yu2022dual}, and CMGAN \cite{cao2022cmgan}) were selected to compare with MP-SENet. 
Six commonly used objective evaluation metrics were chosen to evaluate the enhanced speech quality, including PESQ,  segmental signal-to-noise ratio (SSNR), short-time objective intelligibility (STOI), and three composite measures (CSIG, CBAK, and COVL) which predict the mean opinion score (MOS) on signal distortion, background noise intrusiveness, and overall effect, respectively. For all the metrics, higher values indicate better performance.

\begin{figure}[t]
  \centering
  \includegraphics[width=\linewidth]{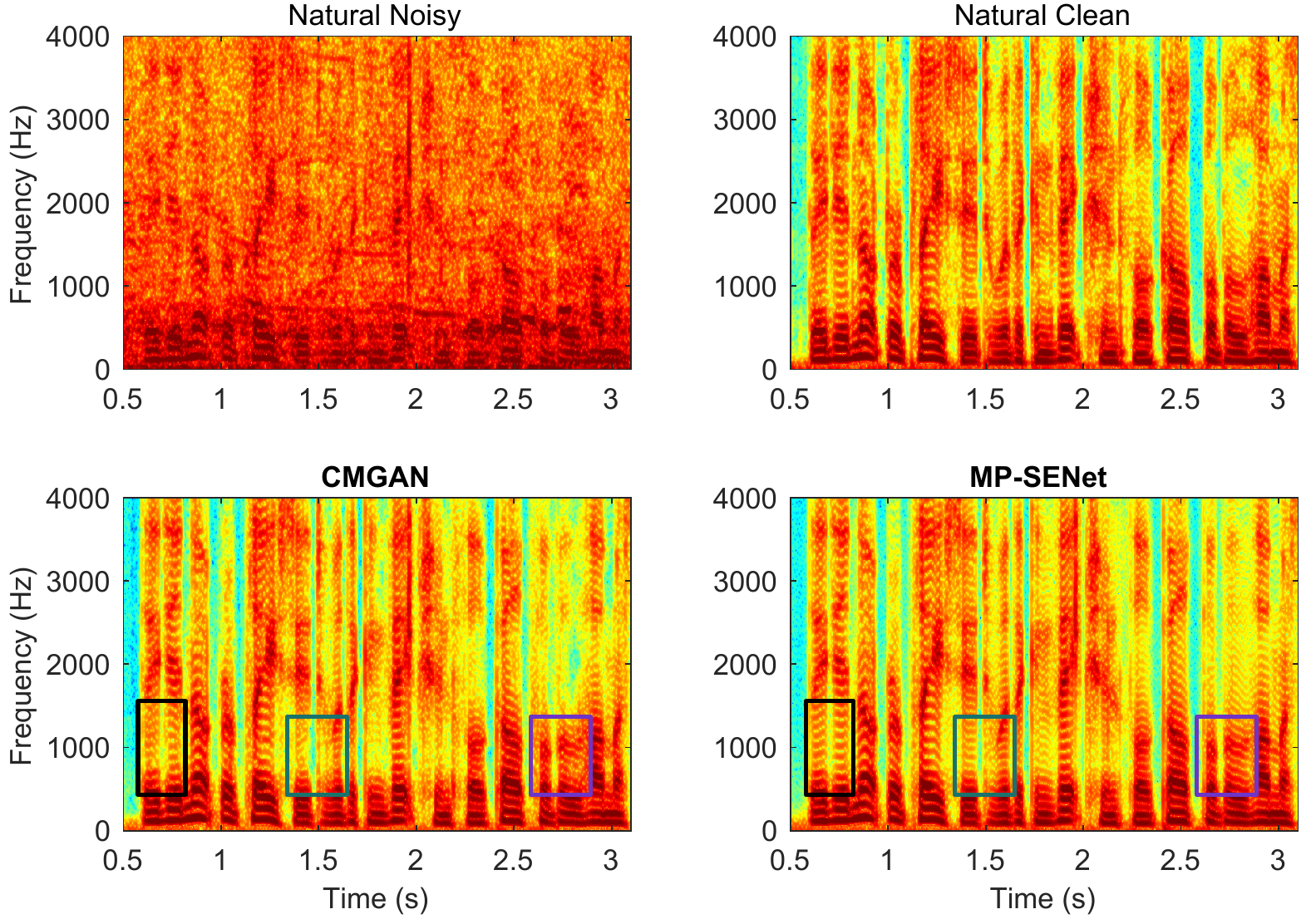}
  \caption{Spectrogram visualization of the natural noisy speech, natural clean speech, and speeches enhanced by the CMGAN and our proposed MP-SENet. For a more intuitive comparison, we only visualised the low-frequency areas.}
  \label{fig: spec}
\end{figure}

The objective results are presented in Table.~\ref{tab: objective evaluation}. Obviously, our proposed MP-SENet achieved a satisfactory PESQ score of 3.50 and outperformed all other SE methods on most metrics, reflecting our model's strong denoising ability. From Table.~\ref{tab: objective evaluation}, we can see that time-domain methods were always inferior to TF-domain methods. Inside the TF-domain methods, only our proposed MP-SENet and PHASEN used both magnitude and phase spectra as input conditions. In spite of that, our proposed MP-SENet had 0.51, 0.52, 0.40, 0.60, and 0.46 improvements on the PESQ, CSIG, CBAK, COVL, and SSNR scores compared to PHASEN. The improvement was significant, proving that the magnitude mask decoder and phase decoder can recover more precise magnitude and phase spectra. When compared to the four TF-domain SOTA approaches, MP-SENet performed the best on the PESQ and three MOS-based metrics but slightly worse than DB-AIAT and CMGAN on the SSNR. For more evidence, we visualized the spectrograms of speeches enhanced by MP-SENet and CMGAN as shown in Fig.~\ref{fig: spec}. By comparing the contents of boxes with the same color, we can see that the MP-SENet alleviated the damage of low-frequency harmonic structures that occurred in CMGAN. This result indicated that predicting the phase directly instead of using complex spectrum refinement can alleviate the compensation effect between magnitude and phase and improve the enhancement precision of the phase spectra. In general, our proposed MP-SENet achieved a new SOTA performance among the objective metrics with a moderate model size of 2.05 M parameters.

\vspace{-1mm}
\subsection{Ablation study}
\vspace{-1mm}

To verify the role of each key component in the MP-SENet, we performed ablation studies and the results are presented in Table.~\ref{tab: ablation}. Obviously, the objective metrics collapsed when removing the magnitude compression operation (``$w/o$ Mag. comp.") and apparently degraded when replacing the LSigmoid activation with PReLU (``$w/o$ LSigmoid"), which demonstrated the magnitude compression operation and LSigmoid were both effective for precise magnitude prediction. When we removed the phase decoder and combined the enhanced magnitude spectrum with the noisy phase spectrum to generate a waveform (``$w/o$ Pha. dec."), all the metrics greatly degraded which proved that phase prediction was indispensable. To further investigate the effects of phase optimization approaches, we conducted ablation studies on the phase loss (``$w/o$ Pha. loss") and complex loss (``$w/o$ Com. loss"), which explicitly and implicitly optimized the phase, respectively. Results demonstrated that although both of them contributed to the overall performance, the explicit phase optimization was quite pivotal to SE tasks. Lastly, we ablated the metric discriminator (``$w/o$ Metric disc.") to assess the generator's ability and improve the training efficiency. Surprisingly, our proposed MP-SENet without discriminator was still capable of high-quality SE, and its performance was even comparable to that of CMGAN.

\begin{table}[t]
  \caption{Results of the ablation studies.}
  \label{tab: ablation}
  \centering
  \resizebox{\linewidth}{!}{
  \begin{tabular}{lccccc}
    \toprule
    \textbf{Method} & \textbf{PESQ} & \textbf{CSIG} & \textbf{CBAK} & \textbf{COVL} & \textbf{SSNR} \\
    \midrule
    MP-SENet       & \textbf{3.50} & \textbf{4.73} & \textbf{3.95} & \textbf{4.22} & \textbf{10.64} \\
    $w/o$ Mag. comp.      & 2.97 & 4.06 & 3.61 & 3.56 & 9.39  \\
    $w/o$ LSigmoid       & 3.40 & 4.67 & 3.87 & 4.14 & 10.10 \\
    $w/o$ Pha. dec.   & 3.31 & 4.61 & 3.82 & 4.06 & 10.05 \\
    $w/o$ Pha. loss 	 & 3.39 & 4.65 & 3.87 & 4.13 & 10.19 \\
    $w/o$ Com. loss 	 & 3.44 & 4.72 & 3.89 & 4.19 & 10.21 \\
    $w/o$ Metric disc.   & 3.39 & 4.69 & 3.89 & 4.15 & 10.48 \\
    \bottomrule
  \end{tabular}}
\end{table}

\section{Conclusions}
In this paper, we proposed an SE model called MP-SENet, which denoised the magnitude and phase spectra in parallel. The overall structure of the MP-SENet was a Conformer-embedded codec architecture. The encoder encoded noisy magnitude and phase spectra, and the parallel magnitude mask decoder and phase decoder decoded out the clean magnitude and phase spectra, respectively. The major breakthrough of the MP-SENet lay in the direct enhancement of phase spectra. Experimental results show that our proposed MP-SENet achieved a SOTA performance on the VoiceBank+DEMAND dataset compared with other advanced SE methods. Moreover, ablation studies verified the effectiveness of each component and optimization method in the MP-SENet. Applying the parallel magnitude and phase enhancement method to other SE tasks (e.g., speech dereverberation, speech separation, and speech super-resolution) will be the focus of our future work.



\bibliographystyle{IEEEtran}
\bibliography{mybib}

\end{document}